\documentclass[axodraw,onecolumn,showpacs,preprintnumbers,amsmath,amssymb]{revtex4}
\usepackage{graphicx}
\usepackage{dcolumn}
\usepackage{bm}
\begin{document}
%%%%%%%%%%%%%%%%%%%%%%%%
\newcommand{\hs}{\hspace*{0.5cm}}
\newcommand{\vs}{\vspace*{0.5cm}}
\newcommand{\be}{\begin{equation}}
\newcommand{\ee}{\end{equation}}
\newcommand{\bea}{\begin{eqnarray}}
\newcommand{\eea}{\end{eqnarray}}
\newcommand{\ben}{\begin{enumerate}}
\newcommand{\een}{\end{enumerate}}
\newcommand{\bde}{\begin{widetext}}
\newcommand{\ede}{\end{widetext}}
\newcommand{\nn}{\nonumber}
\newcommand{\crn}{\nonumber \\}
\newcommand{\non}{\nonumber}
\newcommand{\noi}{\noindent}
\newcommand{\al}{\alpha}
\newcommand{\la}{\lambda}
\newcommand{\bet}{\beta}
\newcommand{\ga}{\gamma}
\newcommand{\va}{\varphi}
\newcommand{\om}{\omega}
\newcommand{\pa}{\partial}
\newcommand{\fr}{\frac}
\newcommand{\bc}{\begin{center}}
\newcommand{\ec}{\end{center}}
\newcommand{\Ga}{\Gamma}
\newcommand{\de}{\delta}
\newcommand{\De}{\Delta}
\newcommand{\ep}{\epsilon}
\newcommand{\varep}{\varepsilon}
\newcommand{\ka}{\kappa}
\newcommand{\La}{\Lambda}
\newcommand{\si}{\sigma}
\newcommand{\Si}{\Sigma}
\newcommand{\ta}{\tau}
\newcommand{\up}{\upsilon}
\newcommand{\Up}{\Upsilon}
\newcommand{\ze}{\zeta}
\newcommand{\ps}{\psi}
\newcommand{\Ps}{\Psi}
\newcommand{\ph}{\phi}
\newcommand{\vph}{\varphi}
\newcommand{\Ph}{\Phi}
\newcommand{\Om}{\Omega}
\def\lappeq{\mathrelc}
%%%%%%%%%%%%%%%%%%%%%%%%

\title{Bilepton contributions to the neutrinoless double beta decay\\
 in the economical 3-3-1 model}

\author{Dang Van Soa}
\email{dvsoa@assoc.iop.vast.ac.vn} \affiliation{Department of
Physics, Hanoi University of Education, Hanoi, Vietnam}

\author{Phung Van Dong}
\email{pvdong@iop.vast.ac.vn}
\author{Trinh Thi Huong}
\email{tthuong@iop.vast.ac.vn}
\author{Hoang Ngoc Long}
\email{hnlong@iop.vast.ac.vn} \affiliation{Institute of Physics,
VAST, P. O. Box 429, Bo Ho, Hanoi 10000, Vietnam}

\date{\today}
\begin{abstract}
Possible contributions of the bilepton to the neutrinoless double
beta $(\beta\beta)_{0\nu}$ decay in the economical 3-3-1 model are
discussed. We show that the $(\beta\beta)_{0\nu}$ decay in this
model is due to both sources---Majorana $\langle M_{\nu}\rangle_{L}$
and Dirac $\langle M_{\nu}\rangle_{D}$ neutrino masses. If the
mixing angle between charged gauge bosons, the standard model $W$
and bilepton $Y$, is in range of the ratio of neutrino masses
$\langle M_{\nu}\rangle_{L}/\langle M_{\nu}\rangle_{D}$, both the
Majorana and Dirac masses simultaneously give contributions dominant
to the decay. As results, constraints on the bilepton mass are also
given.

\end{abstract}


\maketitle

\section{\label{Intro}Introduction}

In the standard model (SM) of strong and electroweak interactions,
the neutrinos are strictly massless due to absence of right-handed
chiral states ($\nu_R$) and requirement of $\mathrm{SU}(2)_L\otimes
\mathrm{U}(1)_Y$ gauge invariance and renormalizability. Recent
experimental results of SuperKamiokande Collaboration~\cite{superK},
KamLAND~\cite{kam} and SNO~\cite{sno} confirm that the neutrinos
have tiny masses and oscillate, this implies that the SM must be
extended. Among beyond-SM extensions, the models based on
$\mathrm{SU}(3)_C\otimes \mathrm{SU}(3)_L \otimes \mathrm{U}(1)_X$
(3-3-1) gauge group \cite{ppf,flt} have some intriguing features:
First, they can give partial explanation of the generation number
problem. Second,  the third quark generation has to be different
from the first two, so this leads to possible explanation of why top
quark is uncharacteristically heavy.

In one of 3-3-1 models three lepton triplets are of the form
$(\nu_L,l_L,\nu^c_R)$ and the scalar sector is minimal with just two
Higgs triplets, hence it has been called the economical 3-3-1
model~\cite{dlns}. The general Higgs sector is very simple and
consists of three physical scalars (two neutral and one charged) and
eight Goldstone bosons---the needed number for massive gauge bosons
\cite{dls}. The model is consistent and possesses key properties:
(i) There are three quite different scales of vacuum expectation
values (VEVs): $u \sim {\cal O}(1) \ \mathrm{GeV}$, $v \approx 246\
\mathrm{GeV}$, and $\om \sim {\cal O}(1)\ \mathrm{TeV}$; (ii) There
exist two types of Yukawa couplings with very different strengths,
the lepton-number conserving (LNC) $h$'s and the lepton-number
violating (LNV) $s$'s, satisfying $ s \ll h$. The resulting model
yields interesting physical phenomenologies due to mixings in the
Higgs \cite{dls}, gauge \cite{dln1} and quark \cite{dhhl} sectors.

Despite present experimental advances in neutrino physics, we have
not yet known if the neutrinos are Dirac or Majorana particles. If
the neutrinos are Majorana ones, the mass terms violate lepton
number by two units, which may result in important consequences in
particle physics and cosmology. A crucial process that will help in
determining neutrino nature is the neutrinoless double beta
$(\beta\beta)_{0\nu}$ decay~\cite{for}. It is also a typical process
which requires violation of the lepton number, although it could say
nothing about the value of the mass. This is because although
right-handed currents and/or scalar bosons may affect the decay
rate, it has been shown that whatever the mechanism of this decay is
a non-vanishing neutrino mass~\cite{sche}. In some models
$(\beta\beta)_{0\nu}$ decay can proceed with arbitrary small
neutrino mass via scalar boson exchange~\cite{plei1}.

The mechanism involving a trilinear interaction of the scalar bosons
was proposed in Ref.~\cite{moha} in the context of model with
$\mathrm{SU}(2)\otimes \mathrm{U}(1)$ symmetry with doublets and a
triplet of scalar bosons. However, since in these types of models
there is no large mass scale~\cite{hax}, the contribution of the
trilinear interaction is, in fact, negligible. In general, in models
with that symmetry, a fine tuning is needed if we want the trilinear
terms to give important contributions to the $(\beta\beta)_{0\nu}$
decay~\cite{esco}. It was shown in Ref.~\cite{foot, mon} that in
3-3-1 models, which has a rich Higgs bosons sector, there are new
many contributions to the $(\beta\beta)_{0\nu}$ decay. In recent
work~\cite{alex}, authors showed that the implementation of
spontaneous breaking of the lepton number in the 3-3-1 model with
right-handed neutrinos gives rise to fast neutrino decay with
Majoron emission and generates a bunch of new contributions to the
$(\beta\beta)_{0\nu}$ decay.

In an earlier work~\cite{dls1} we have analyzed the neutrino masses
in the economical 3-3-1 model. The masses of neutrinos are given by
three different sources widely ranging over the mass scales
including the GUT's and the small VEV $u$ of spontaneous lepton
breaking. With a finite renormalization in mass, the spectrum of
neutrino masses is neat and can fit the data. In this work, we will
discuss possible contributions of the bilepton to the
$(\beta\beta)_{0\nu}$ decay in the considering model. We show that
in contradiction with previous analysis, the $(\beta\beta)_{0\nu}$
decay arises from two different sources, which require both the
non-vanishing Majorana and Dirac neutrino masses. If the mixing
angle between the charged gauge bosons is in range of the ratio of
neutrino masses  $\langle M_{\nu}\rangle_{L}/\langle
M_{\nu}\rangle_{D}$, both the Majorana and Dirac masses
simultaneously give dominant contributions to the decay. The
constraints on the bilepton mass are also given.

The rest of this paper is organized as follows: In section
\ref{model} we give a brief review of the economical 3-3-1 model.
Charged currents and a new bound of the mixing angle are given in
section \ref{mixing}. Section \ref{0n2betadecays} is devoted to
detailed analysis of the possible contributions of the bilepton to
the $(\beta\beta)_{0\nu}$ decay.
We summarize our results and make conclusions in the last section - Sec. \ref{conclus}.\\

\section{\label{model} A review of the model }

The particle content in this model which is anomaly free is given
as follows  \cite{dlns}\bea \psi_{aL} &=& \left(
               \nu_{aL}, l_{aL}, (\nu_{aR})^c
\right)^T \sim (3, -1/3),\hs l_{aR}\sim (1, -1),\hs a = 1, 2, 3,
\crn
 Q_{1L}&=&\left( u_{1L},  d_{1L}, U_L \right)^T\sim
 \left(3,1/3\right),\hs Q_{\al L}=\left(
  d_{\al L},  -u_{\al L},  D_{\al L}
\right)^T\sim (3^*,0),\hs \al=2,3,\crn u_{a
R}&\sim&\left(1,2/3\right),\hs d_{a R} \sim
\left(1,-1/3\right),\hs U_{R}\sim \left(1,2/3\right),\hs D_{\al R}
\sim \left(1,-1/3\right),\eea where the values in the parentheses
denote quantum numbers based on the
$\left(\mbox{SU}(3)_L,\mbox{U}(1)_X\right)$ symmetry. Unlike the
usual 3-3-1 model with right-handed neutrinos, where the third
family of quarks should be discriminating, in the model under
consideration the {\it first} family has to be different from the
two others \cite{dhhl}. The electric charge operator in this case
takes a form\be Q=T_3-\fr{1}{\sqrt{3}}T_8+X,\label{eco}\ee where
$T_i$ $(i=1,2,...,8)$ and $X$, respectively, stand for
$\mbox{SU}(3)_L$ and $\mbox{U}(1)_X$ charges. The electric charges
of the exotic quarks $U$ and $D_\al$ are the same as of the usual
quarks, i.e., $q_{U}=2/3$, $q_{D_\al}=-1/3$.

The spontaneous symmetry breaking in this model is obtained by two
stages: \be \mathrm{SU}(3)_L\otimes \mathrm{U}(1)_X \rightarrow
\mathrm{SU}(2)_L\otimes\mathrm{U}(1)_Y \rightarrow
\mathrm{U}(1)_Q.\ee The first stage is achieved by a Higgs scalar
triplet with a VEV given by \bea \chi=\left(\chi^0_1, \chi^-_2,
\chi^0_3 \right)^T \sim \left(3,-1/3\right),\hs
\langle\chi\rangle=\fr{1}{\sqrt{2}}\left(u, 0, \om
\right)^T.\label{vevc}\eea The last stage is achieved by another
Higgs scalar triplet needed with the VEV as follows \bea
\phi=\left(\phi^+_1, \phi^0_2, \phi^+_3\right)^T \sim
\left(3,2/3\right),\hs
\langle\phi\rangle=\fr{1}{\sqrt{2}}\left(0,v,0
\right)^T.\label{vevp}\eea

The Yukawa interactions which induce masses for the fermions can
be written in the most general form: \be {\mathcal
L}_{\mathrm{Y}}={\mathcal L}_{\mathrm{LNC}} +{\mathcal
L}_{\mathrm{LNV}},\ee in which, each part is defined by \bea
{\mathcal L}_{\mathrm{LNC}}&=&h^U\bar{Q}_{1L}\chi
U_{R}+h^D_{\al\beta}\bar{Q}_{\al L}\chi^* D_{\beta R}\crn
&&+h^l_{ab}\bar{\psi}_{aL}\phi
l_{bR}+h^\nu_{ab}\ep_{pmn}(\bar{\psi}^c_{aL})_p(\psi_{bL})_m(\phi)_n
\crn && +h^d_{a}\bar{Q}_{1 L}\phi d_{a R}+h^u_{\al a}\bar{Q}_{\al
L}\phi^* u_{aR}+ H.c.,\label{y1}\\ {\mathcal
L}_{\mathrm{LNV}}&=&s^u_{a}\bar{Q}_{1L}\chi u_{aR}+s^d_{\al
a}\bar{Q}_{\al L}\chi^* d_{a R}\crn && +s^D_{ \al}\bar{Q}_{1L}\phi
D_{\al R}+s^U_{\al }\bar{Q}_{\al L}\phi^* U_{R}+
H.c.,\label{y2}\eea where $p$, $m$ and $n$ stand for
$\mathrm{SU}(3)_L$ indices.

The VEV $\om$ gives mass for the exotic quarks $U$, $D_\al$ and the
new gauge bosons $Z^{\prime},\ X,\ Y$, while the VEVs $u$ and $v$
give mass for all the ordinary fermions and gauge bosons
\cite{dhhl,dls1}. To keep a consistency with the effective theory,
the VEVs in this model have to satisfy the constraint \be u^2 \ll
v^2 \ll \om^2. \label{vevcons} \ee
 In addition we can derive $v\approx
v_{\mathrm{weak}}=246\ \mbox{GeV}$ and $|u| \leq2.46\ \mbox{GeV}$
from the mass of $W$ boson and the $\rho$ parameter \cite{dlns},
respectively. From atomic parity violation in cesium, the bound for
the mass of new natural gauge boson is given by $M_{Z^{\prime}}>564
\ \mbox{GeV}$ $(\om > 1400\ \mbox{GeV})$ \cite{dln1}. From the
analysis on quark masses, higher values for $\om$ can be required,
for example, up to $10\ \mbox{TeV}$ \cite{dhhl}.

The Yukawa couplings of (\ref{y1}) possess an extra global symmetry
\cite{changlong} not broken by $ v, \omega$ but by $u$. From these
couplings, one can find the following lepton symmetry $L$ as in
Table \ref{lnumber} (only the fields with nonzero $L$ are listed;
all other ones have vanishing $L$).
\begin{table}[h]
 \caption{\label{lnumber} Nonzero lepton number $L$
 of the model particles.}
\begin{ruledtabular}
\begin{tabular}{lcccccccc}
  Field
&$\nu_{aL}$&$l_{aL,R}$&$\nu^c_{aR}$ & $\chi^0_1$&$\chi^-_2$ &
$\phi^+_3$ & $U_{L,R}$ & $D_{\alpha L,R}$\\
    \hline \\
        $L$ & $1$ & $1$ & $-1$ & $2$&$2$&$-2$&$-2$&$2$
\end{tabular}
 \end{ruledtabular}
\end{table} Here
$L$ is broken by $u$ which is behind $L(\chi^0_1)=2$, i.e., $u$ {\it
is a kind of the SLB scale} \cite{major-models}.

It is interesting that the exotic quarks also carry the lepton
number; therefore, this $L$ obviously does not commute with the
gauge symmetry. One can then construct a new conserved charge $\cal
L$ through $L$ by making a linear combination $L= xT_3 + yT_8 +
{\cal L} I$. Applying $L$ on a lepton triplet, the coefficients will
be determined \be L = \fr{4}{\sqrt{3}}T_8 + {\cal L} I
\label{lepn}.\ee Another useful conserved charge $\cal B$ exactly
not broken by $u$, $v$ and $\om$ is usual baryon number $B ={\cal B}
I$. Both the charges $\mathcal{L}$ and $\mathcal{B}$ for the fermion
and Higgs multiplets are listed in Table~\ref{bcharge}.
\begin{table}[h]
\caption{\label{bcharge} ${\cal B}$ and ${\cal L}$ charges of the
model multiplets.} \begin{ruledtabular}
\begin{tabular}{lcccccccccc}
 Multiplet & $\chi$ & $\phi$ & $Q_{1L}$ & $Q_{\al L}$ &
$u_{aR}$&$d_{aR}$ &$U_R$ & $D_{\al R}$ & $\psi_{aL}$ & $l_{aR}$
\\ \hline \\ $\cal B$-charge &$0$ & $ 0  $ &  $\fr 1 3  $ & $\fr 1 3
$& $\fr 1 3  $ &
 $\fr 1 3  $ &  $\fr 1 3  $&  $\fr 1 3  $&
 $0  $& $0$ \\ \hline \\
 $\cal L$-charge &$\fr 4 3$ & $-\fr 2 3  $ &
   $-\fr 2 3  $ & $\fr 2 3  $& 0 & 0 & $-2$& $2$&
 $\fr 1 3  $& $ 1   $
\end{tabular}
 \end{ruledtabular}
\end{table}

Let us note that the Yukawa couplings of (\ref{y2}) conserve
$\mathcal{B}$, however, violate ${\mathcal L}$ with $\pm 2$ units
which implies that these interactions are much smaller than the
first ones \cite{dhhl}: \be s_a^u, \ s_{\al a}^d,\ s_\al^D, \
s_\al^U \ll h^U,\ h_{\al \bet}^D,\ h_a^d,\ h_{\al
a}^u.\label{dkhsyu}\ee

\section{\label{mixing}Charged currents and a new bound of the mixing angle}

A consequence of $u\neq 0$ is that the SM gauge boson $W'$ and
bilepton $Y'$ mix \bea {\cal
L}^{\mathrm{CG}}_{\mathrm{mass}}=\fr{g^2}{4}(W'^-,Y'^-)\left(%
\begin{array}{cc}
  u^2+v^2 & u\om \\
  u\om & \om^2+v^2 \\
\end{array}%
\right)\left(%
\begin{array}{c}
  W'^{+} \\
  Y'^{+} \\
\end{array}%
\right).\nn\eea Physical charged gauge bosons are given by
 \bea W &=& \cos\theta\
W'+\sin\theta\ Y',\crn Y &=& -\sin\theta\ W'+\cos\theta\
Y',\label{xy}\eea where the mixing angle is \be \tan\theta=\fr{u}{\om}.\label{theta}\ee\\
There exist LNV terms in the charged currents proportional to $\sin
\theta$ \bea H^{\mathrm{CC}}=\fr{g}{\sqrt{2}}\left(J^{\mu+}_W
W^-_\mu + J^{\mu+}_Y Y^-_\mu + H.c.\right), \eea with \bea
J^{\mu+}_W&=&c_\theta \left(\overline{l}_{aL}\ga^\mu
\nu_{aL}+\overline{d}_{aL}\ga^\mu u_{aL}\right)\crn &&-s_\theta
\left(\overline{l}_{aL}\ga^\mu
\nu^c_{aR}+\overline{d}_{1L}\ga^\mu U_{L}+\overline{D}_{\al L}\ga^\mu u_{\al L}\right),\label{dongw}\\
J^{\mu+}_Y&=&c_\theta \left(\overline{l}_{aL}\ga^\mu
\nu^c_{aR}+\overline{d}_{1L}\ga^\mu U_{L}+\overline{D}_{\al
L}\ga^\mu u_{\al L}\right)\crn &&+s_\theta \left(\overline{l}_{aL}
\ga^\mu\nu_{aL}+\overline{d}_{a L}\ga^\mu u_{a
L}\right).\label{dongy}\eea

As in Ref. \cite{dlns}, the constraint on the $W-Y$ mixing angle
$\theta$ from the $W$ width is given by $\sin \theta \leq 0.08$.
However, in the following we will show that a more stricter bound
can obtain from the invisible $Z$ width through the {\it unnormal
neutral current} of LNV  \bea {\mathcal L}^{\mathrm{NC}
}_{\mathrm{unnormal}}&=&-\fr{g
 t_{2\theta}g_{kV}(\nu)}{ c_W}\left(\overline{\nu}_{aL}\ga^\mu
 \nu^c_{aR}+\overline{u}_{1L}\ga^\mu U_{L}\right.\crn
&&\left.-\overline{D}_{\al L}\ga^\mu d_{\al L}\right) Z^k_\mu + H.c.
\label{un}, \eea where the neutrino coupling constants ($g_{kV},
k=1,2$) are given by
 \bea g_{1V}
(\nu_L) &\simeq& \fr{c_\va -s_\va \sqrt{4c^2_W-1}}{2}, \eea \bea
g_{2V} (\nu_L) &\simeq& \fr{s_\va +c_\va
\sqrt{4c^2_W-1}}{2}\label{un1}. \eea Let us note that the LNV
interactions mediated by neutral gauge bosons $Z^1$ and $Z^2$ exist
only in the neutrino and exotic quark sectors. The interactions in
(\ref{un}) for the neutrinos lead to additional invisible-decay
modes to the $Z$ boson. For each generation of lepton, due to the
angle $\va$ has to be very small \cite{dln1,guti}, the corresponding
invisible-decay width gets approximation
 \bea \Ga_{\nu_{L}
N_{L}}&\simeq&\frac{1}{2}t^2_{2\theta}\Big( 1 +
\mathcal{O}(s^2_{\va})\Big)\Ga^{SM}_{\nu\overline{\nu}}\label{un2},\eea
where $N_{L}=\nu^c_{aR}$ and
$\Ga^{\mathrm{SM}}_{\nu\overline{\nu}}=\fr{G_F
M^3_Z}{12\pi\sqrt{2}}$ is the SM prediction for the decay rate of
$Z$ into a pair of neutrinos. The experimental data for the total
invisible neutrino decay modes give us~\cite{pdg} \bea
\Ga^{\mathrm{exp}}_{\mathrm{invi}}&=&(2.994\pm
0.012)\Ga^{\mathrm{SM}}_{\nu\overline{\nu}}.\label{un4}\eea From
(\ref{un2}) and (\ref{un4}) we get an upper limit for the mixing
angle \bea t_{\theta}\leq 0.03,\label{un5}\eea {\it which is smaller
than that given in Ref. \cite{dlns}.}\\

Let us briefly discuss the neutrino mass. The masses of neutrinos in
this model are given by three different sources widely ranging over
the mass scales including the GUT's and the small VEV $u$ of
spontaneous lepton breaking. At the tree-level, the model contains
three Dirac neutrinos: one massless, two large with degenerate
masses in the range of the electron mass. At the one-loop level, the
left-handed and right-handed neutrinos obtain Majorana masses
$M_{L,R}$ in orders of $10^{-2}-10^{-3}\ \mathrm{eV}$ and degenerate
in $M_R=-M_L$, while the Dirac masses get a large reduction down to
$\mathrm{eV}$ scale through a finite mass renormalization. In this
model, the contributions of new physics are strongly signified, the
degenerations in the masses and the last hierarchy between the
Majorana and Dirac masses can be completely removed by heavy
particles. All the neutrinos get mass and can fit the data (for
details, see Ref.~\cite{dls1}).

\section{\label{0n2betadecays}Bilepton contributions to the neutrinoless double beta decay}
The $(\beta\beta)_{0\nu}$ decay is the typical process which
requires violation of the lepton number, thus it can be useful in
probing new physics beyond the standard model. The interactions that
lead to the $(\beta\beta)_{0\nu}$ decay involve hadrons and leptons.
For the case of the standard contribution, its amplitude can be
written as~\cite{alex} \bea M_{(\beta\beta)_{0\nu}} =
\frac{g^4}{4m^4_{W}}
M^h_{\mu\nu}\overline{u}\gamma^{\mu}P_L\frac{q\!\!\!/+ m_{\nu}}{q^2
- m^2_{\nu}}\gamma^{\nu} P_R v\label{m1},\eea with $M^h_{\mu\nu}$
carrying the hadronic information of the process and $P_{R,L}=
\frac{(1\pm\gamma_5)}{2}$. In the presence of neutrino mixing and
considering that $m^2_{\nu}\ll q^2$, we can write \bea
M_{(\beta\beta)_{0\nu}} = A_{(\beta\beta)_{0\nu}}
M^h_{\mu\nu}\overline{u}P_R\gamma^{\mu}\gamma^{\nu} v,\label{m2}\eea
where \bea A_{(\beta\beta)_{0\nu}} = \frac{g^4\langle
M_{\nu}\rangle}{4m^4_{W}\langle q^2\rangle}\label{m3}\eea is the
strength of effective coupling of the standard contribution. For the
case of three neutrino species
 $\langle M_{\nu}\rangle = \sum U^2_{ei}m_{\nu i}$ is the effective
 neutrino mass and  $\langle q^2\rangle$
is the average of the transferred squared four-momentum.

The contributions to the $(\beta\beta)_{0\nu}$ decay in our model
coming from the charged gauge bosons $W^-$ and $Y^-$ dominate the
process. As the $(\beta\beta)_{0\nu}$ decay has not been
experimentally detected yet, the analysis we do here is to obtain  a
new contributions and to compare them with the standard
one~\cite{sche, mon}. Feynman diagrams for contributions are
depicted in the figures: Fig.1, Fig.2 and Fig.3, respectively.
Left-handed figures {\bf(a)} are given by the non - vanishing
Majorana mass, the right-handed figures {\bf(b)} -- Dirac
mass.\\

For the standard contribution as depicted in Fig.(1.a), its
effective coupling takes the form \bea A_{(\beta\beta)_{0\nu}}(1.a)
= \frac{g^4\langle M_{\nu}\rangle_{L}}{4m^4_{W}\langle
q^2\rangle}c^4_{\theta},\label{a2}\eea where $M_L$ is the Majorana
mass. The first new contribution involves only $W^-$ as of the
standard one, but now interacts with two charged currents $J_{\mu}$
and $J^c_\mu$ as depicted in Fig.(1.b). It is to be noted that in
this case the Dirac mass gives the contribution to the effective
coupling \bea A_{(\beta\beta)_{0\nu}}(1.b) = \frac{g^4\langle
M_{\nu}\rangle_{D}}{4m^4_{W}\langle
q^2\rangle}c^3_{\theta}s_{\theta},\label{a3}\eea  where $M_D$ is the
Dirac mass.

From Eqs. (\ref{a2}) and (\ref{a3}) we see that the LNV in the
$(\beta\beta)_{0\nu}$ decay arises from two different sources
identified by the non-vanishing Majorana and Dirac mass terms,
respectively. In Fig.(1.a) the LNV is due to the Majorana mass,
while that in Fig.(1.b) is by the LNV coupling of $W$ boson to the
charged current (the term is proportional to $\sin \theta$). In
comparing both effective couplings, we obtain the ratio \bea
\frac{A_{(\beta\beta)_{0\nu}}(1.b) }{A_{(\beta\beta)_{0\nu}}(1.a) }=
\frac{\langle M_{\nu}\rangle_{D}}{\langle
M_{\nu}\rangle_{L}}\tan\theta\label{a33}.\eea From (\ref{a33}) we
see that the relevance of this contribution depends on  angle
$\theta$ and also the ratio between $\langle M_{\nu}\rangle_{D}$ and
$\langle M_{\nu}\rangle_{L}$. It is worth noting that if $ \langle
M_{\nu}\rangle_{D}.t_{\theta}\sim \langle M_{\nu}\rangle_{L}$ then
both Majorana and Dirac masses simultaneously give  the dominant
contributions to the $(\beta\beta)_{0\nu}$ decay.

Next, we consider contributions that involve both $W^-$ and $Y^-$.
It involves the two currents $J_{\mu}$ and $J^c_\mu$ interacting
with $W$ and $Y$, as depicted in Fig.(2.a) for $\langle
M_{\nu}\rangle_{L}$ and Fig.(2.b) for $\langle
M_{\nu}\rangle_{D}$. The effective couplings in this case are \bea
A_{(\beta\beta)_{0\nu}}(2.a) = \frac{g^4\langle M_{\nu}\rangle_{L}
c^2_{\theta}s^2_{\theta}}{4m^2_{W}m^2_{Y}\langle
q^2\rangle},\label{a4}\eea  and \bea A_{(\beta\beta)_{0\nu}}(2.b)
= \frac{g^4\langle M_{\nu}\rangle_{D}
c^3_{\theta}s_{\theta}}{4m^2_{W}m^2_{Y}\langle
q^2\rangle}.\label{a5}\eea From (\ref{a4}) and (\ref{a5}) we see
that the case with the Majorana mass gives the contribution to the
$(\beta\beta)_{0\nu}$ much smaller than the Dirac one. Comparing
with the standard effective coupling, we get the ratios \bea
\frac{A_{(\beta\beta)_{0\nu}}(2.b) }{A_{(\beta\beta)_{0\nu}}(1.a)
}= \Big(\frac{m^2_{W}}{m^2_{Y}}\Big)\frac{\langle
M_{\nu}\rangle_{D}}{\langle
M_{\nu}\rangle_{L}}\tan\theta,\label{a6}\eea and \bea
\frac{A_{(\beta\beta)_{0\nu}}(2.a) }{A_{(\beta\beta)_{0\nu}}(1.a)
}= \Big(\frac{m^2_{W}}{m^2_{Y}}\Big)\tan^2\theta.\label{a66}\eea
Differing from the previous case, Eq.(\ref{a6}) shows that the
relevance of these contributions depends on the angle $\theta$,
the ratio $\frac{\langle M_{\nu}\rangle_{D}}{\langle
M_{\nu}\rangle_{L}}$ and the bilepton mass also. Suppose that the
new contributions are smaller than the standard one, from Eq.
(\ref{a6}) we get a lower bound for the bilepton mass \bea
m^2_{Y}> m^2_{W}\frac{\langle M_{\nu}\rangle_{D}}{\langle
M_{\nu}\rangle_{L}}\tan\theta. \label{a666}\eea Taking $m^2_{W}=
80.425 GeV, t_{\theta}= 0.03$, the low bounds of mass $m_Y$ in
range of $\frac{\langle M_{\nu}\rangle_{D}}{\langle
M_{\nu}\rangle_{L}}\sim 10^2 - 10^3$~\cite{dls1} are given in
Table III.
\begin{table}[h]
\caption{ The low bound of bilepton mass in range of
$\frac{\langle M_{\nu}\rangle_{D}}{\langle M_{\nu}\rangle_{L}}$ }
\begin{ruledtabular}
\begin{tabular}{lcccccc}
 $\frac{\langle M_{\nu}\rangle_{D}}{\langle M_{\nu}\rangle_{L}}$  & $100$ & $200$ & $400$ & $600$ & $800$&$1000$\\
 \hline \\ $m_{Y}$ ($ GeV$) & $139.0$ & $197.0$ & $278.6$ & $341.2$& $394.0$& $440.5$ \\
\end{tabular}
 \end{ruledtabular}
\end{table}
 It is interesting to note that from
"wrong " muon decay experiments one obtains a bound for the bilepton
mass, $m_{Y}\geq 230 GeV$~\cite{dlns,stl} and the stronger mass
bound has been derived from consideration of an experimental limit
on lepton number violating charged lepton decays~\cite{tull} of
$440$ GeV.\\

From Eq. (\ref{a66}) we see that the order of contribution is much
smaller than standard contribution, this is due to the LNV in the
$(\beta\beta)_{0\nu}$ decay arising from the Majorana mass term and
also the LNV coupling between the bilepton  $Y$ and the charged
current $J^{\mu}$ of ordinary quarks and leptons. Taking $m_{Y}=139$
GeV we obtain \bea \frac{A_{(\beta\beta)_{0\nu}}(2.a)}
{A_{(\beta\beta)_{0\nu}}(1.a)} \leq 3.0\times 10^{-4}. \eea

Now we examine the next four contributions which involve only the
bileptons $Y$. In Fig.(3.a) we display and example of this kind of
contribution where the current $J^c_\mu$ appears in the two
vertices. The effective coupling is \bea
A_{(\beta\beta)_{0\nu}}(3.a) = \frac{g^4\langle M_{\nu}\rangle_{L}
s^4_{\theta}}{4m^4_{Y}\langle q^2\rangle}.\label{a7}\eea For
another case we have also \bea A_{(\beta\beta)_{0\nu}}(3.b) =
\frac{g^4\langle
M_{\nu}\rangle_{D}c_{\theta}s^3_{\theta}}{4m^4_{Y}\langle
q^2\rangle}.\label{a7}\eea Comparing with the standard effective
coupling, we get  \bea \frac{A_{(\beta\beta)_{0\nu}}(3.a)
}{A_{(\beta\beta)_{0\nu}}(1.a) }=
\Big(\frac{m_{W}}{m_{Y}}\Big)^4\tan^4\theta.\label{a8}\eea Using
the above data, the ratio gets an upper limit
  \bea
\frac{A_{(\beta\beta)_{0\nu}}(3.a)
}{A_{(\beta\beta)_{0\nu}}(1.a)}\leq 9.0\times 10^{-8},\label{a9}\eea
which is very small. It is easy to check that the remaining
contributions are much smaller than those with the charged $W$
bosons. This is due to the fact that all the couplings of the
bilepton with ordinary quarks and leptons in the diagrams of Fig.(3)
are LNV.

\section{\label{conclus} Conclusion}
In this paper we have investigated the implications of spontaneous
breaking of the lepton number in the  economical 3-3-1 model in the
$(\beta\beta)_{0\nu}$ decay. We have performed a systematic analysis
of the couplings of all possible contributions of charged gauge
bosons to the decay. The result shows that, the
$(\beta\beta)_{0\nu}$ decay mechanism in the economical 3-3-1 model
requires both the non-vanishing Majorana and Dirac masses. If the
mixing angle between the charged gauge boson and bilepton is in
range of the ratio of neutrino masses $\langle M_{\nu}\rangle_{L}$
and $\langle M_{\nu}\rangle_{D}$ then both the Majorana and Dirac
masses simultaneously give the dominant contributions to the decay.
Basing on the result, the constraints on the bilepton mass are
given. It is interesting to note that the relevance of the new
contributions are dictated by the mixing angle $\theta$, the
effective mass of neutrino and the bilepton mass. By estimating the
order of magnitude of the new contributions, we predicted that the
most robust one is that depicted in Fig.2 whose order of magnitude
is $5.5\times 10^{-4}$ of the standard contribution.\\

Finally, we emphasize that in the considered model, the charged
Higgs boson is the a scalar bilepton (with lepton number $L=\pm 2$)
. Therefore, Yukawa couplings of them with ordinary quarks and
leptons are LNV and very weak (for details, see Ref.~\cite{stth}).
It means that, possible contributions of them to the
$(\beta\beta)_{0\nu}$ decay have to be much smaller than that from
charged gauge bosons.

\section{\label{Acknonw} Acknowledgement}
 One of the authors (D.V. S.) expresses his sincere gratitude
 to the National Center for Theoretical Sciences
of the National Science council of the Republic of China for
financial support. He is also grateful to Prof. Cheng-Wei Chiang and
members of the Department of Physics, National Central University
for warm hospitality during his visit. This work was supported in
part by the National Council for the Natural Sciences of Vietnam.

\newpage
\begin{figure}
\includegraphics{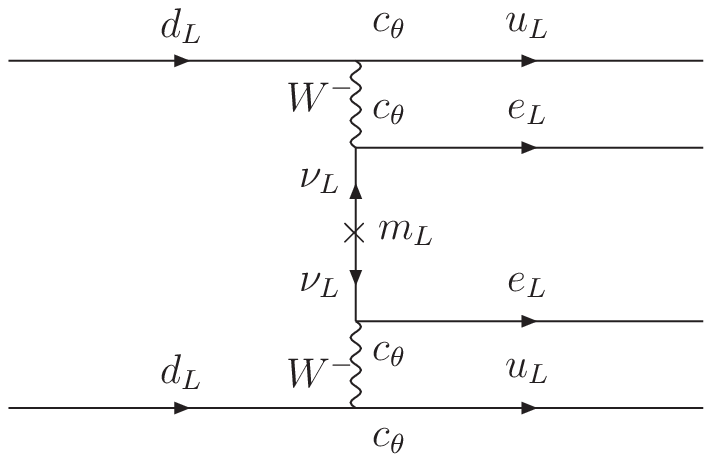}
\includegraphics{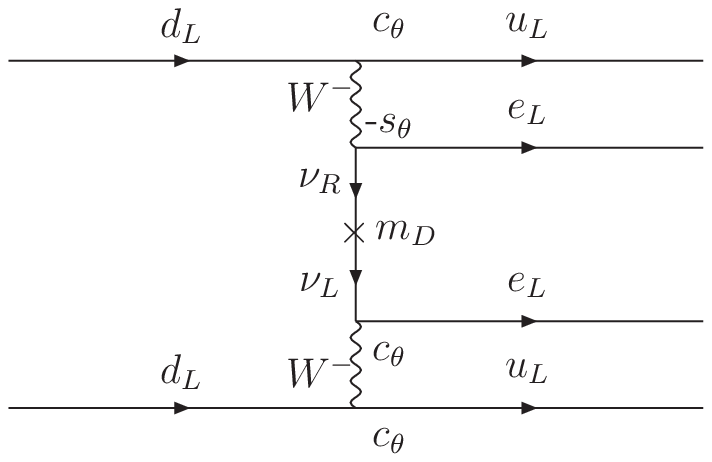}
\caption{\label{W gauge}Contribution of the SM bosons W to the
$(\beta\beta)_{0\nu}$ decay.}
\end{figure}

\begin{figure}
 \includegraphics{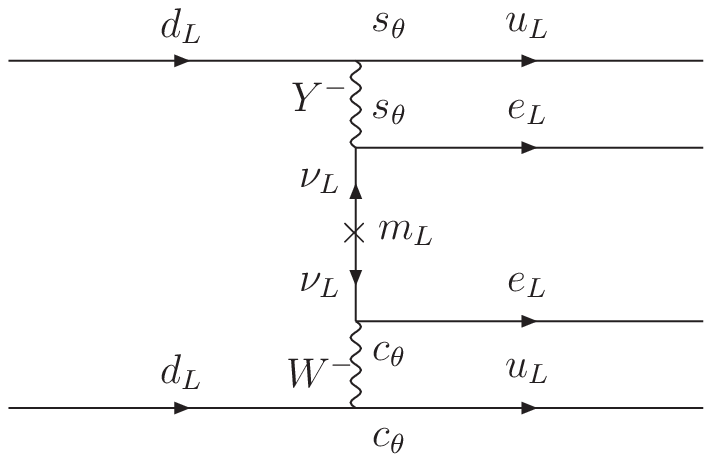}
\includegraphics{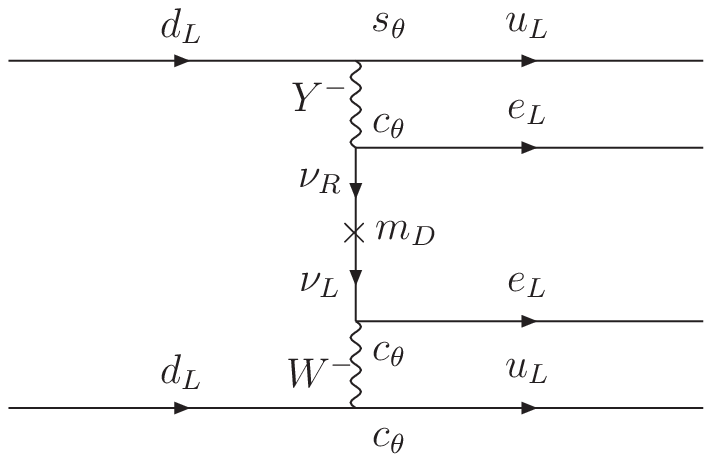}
\caption{\label{WY gauge} Associated contribution of the boson W and
bilepton Y to the $(\beta\beta)_{0\nu}$ decay.}
\end{figure}

 \begin{figure}
 \includegraphics{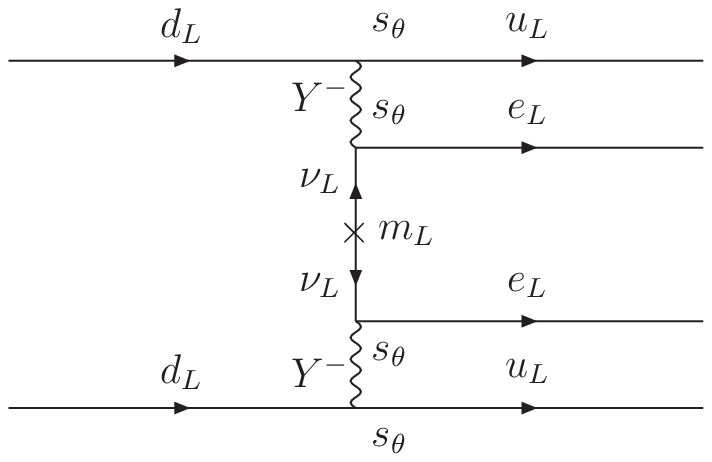}
\includegraphics{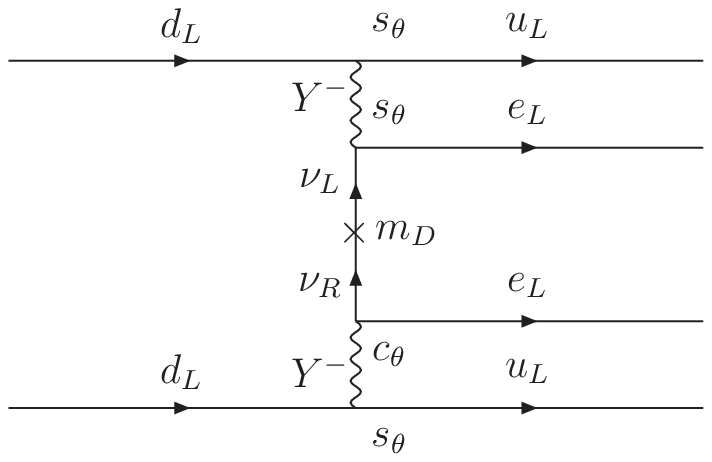}
\caption{\label{Y gauge}Contribution of the bileptons Y to the
$(\beta\beta)_{0\nu}$ decay.}
\end{figure}

\end{document}